\documentclass[aps,pra,preprint,groupedaddress,showpacs]{revtex4-1}

\usepackage{graphicx}
\usepackage{dcolumn}
\usepackage{bm}
\usepackage{amssymb}
\usepackage{mathrsfs}

\begin{document}

\title{Balanced-heterodyne detection of sub-shot-noise optical signals}


\author{Sheng Feng}
\email[]{fengsf2a@mail.hust.edu.cn}
\author{Zehuan Lu}
\author{Jie Zhang}
\author{Chenggang Shao}
\affiliation{School of Physics, Huazhong University of Science and Technology, Wuhan 430074, People's Republic of China}


\date{\today}

\begin{abstract}
As part of an effort to make use of single-mode squeezing for detection of sub-shot-noise optical signals, we study the balanced-heterodyne scheme, for which the corresponding spectral density of the photocurrent fluctuations produced at the output of the detector is calculated as the Fourier transform of their autocorrelation function. We show that, for maximal signal-to-noise ratio enhancement by use of squeezed states of light, an optical signal to be measured in this scheme must be carried in the squeezed quadrature of the carrier field. Most importantly, our analysis indicates that ``the additional heterodyne noise'' can be eliminated in balanced-heterodyne detection for single-mode squeezing under some experimental conditions and, from this, we discuss how this scheme may be exploited in gravitational-wave searching. To demonstrate its practical feasibility, we propose and study a phase-locking technique. 
\end{abstract}


\maketitle

\section{Introduction}
Of the essence is to sense extremely-weak optical signals in several ongoing precision experiments on fundamental physics \cite{ligo1992, fiore2002} and for practical applications \cite{bollinger1996}, in which the shot noise of light resulting from zero-point vacuum fluctuations is becoming a major limiting factor at the quantum level in further improvement of sensitivity. 
To suppress the shot noise, one can resort to the exotic properties of the squeezed states or the entangled states of light \cite{caves1981, walls1983}. As a matter of fact, large degree of squeezing of more than 10dB has been achieved experimentally \cite{sq10}. Depending on the nature of the optical beam to be measured, detection of optical squeezing may be carried out with different schemes. Direct detection is the simplest one and suited for bright amplitude-squeezed light, whereas homodyne detection scheme for quadrature-squeezed light. Another scheme using a cavity to rotate the quadrature orientation of optical sidebands in phase space relative to that of the carrier \cite{levenson1985,shelby1986} can be used to measure the quadrature squeezing of bright light. The practical problem with these three schemes is the intensity noise of light, which is classical and can be substantially suppressed in balanced-homodyne detection \cite{yuen1983}. The great similarity of these schemes is their relying on the detection of the beats between a strong optical mode, the local oscillator or the carrier, and noise sidebands that are correlated with each other and symmetrically located around the strong mode in frequency spectrum. 

Although all the aforesaid detection schemes are playing crucial roles in experimental study for quantum information processing, their applications to precision measurements for detecting sub-shot-noise optical signals are restrained. This is because the beat note between a slowly-varying sub-shot-noise optical signal and the local oscillator is a signal close to DC and will be inevitably contaminated by the low-frequency dark-current noise of a detector. Actually, to circumvent the low-frequency electronic noise, optical squeezing is usually measured at frequencies off the carrier's optical frequency in homodyne detection \cite{sq10}. Apparently, this strategy does not apply to the case where optical squeezing must be measured together with an optical signal that is being under investigation. To shift the frequency of the beat note beyond the low-frequency noise regime of the detector, one may consider the traditional heterodyne scheme with single local oscillator, which unfortunately does not work for detection of squeezed light as proven, for example, by Collett {\it et al.} in \cite{collett1987}. Heterodyne detection schemes for gravitational-wave observation \cite{gea1987, niebauer1991, meers1991, buonanno2003} have been investigated, showing that an additional quantum-noise contribution may exist due to vacuum fluctuations in frequency bands that are twice the heterodyne frequency away from the carrier. In the present paper, we investigate a balanced-heterodyne scheme for detecting sub-shot-noise optical signals, in which the ``additional heterodyne noise'' disappears in certain experimental circumstances, as one will see in the following. One should note that the scenario here is different from that of the theoretical work for two-mode squeezing detection in \cite{marino2007}, which does not deal with the detection of single-mode squeezing.

\begin{figure}[B]
\includegraphics[scale=0.5]{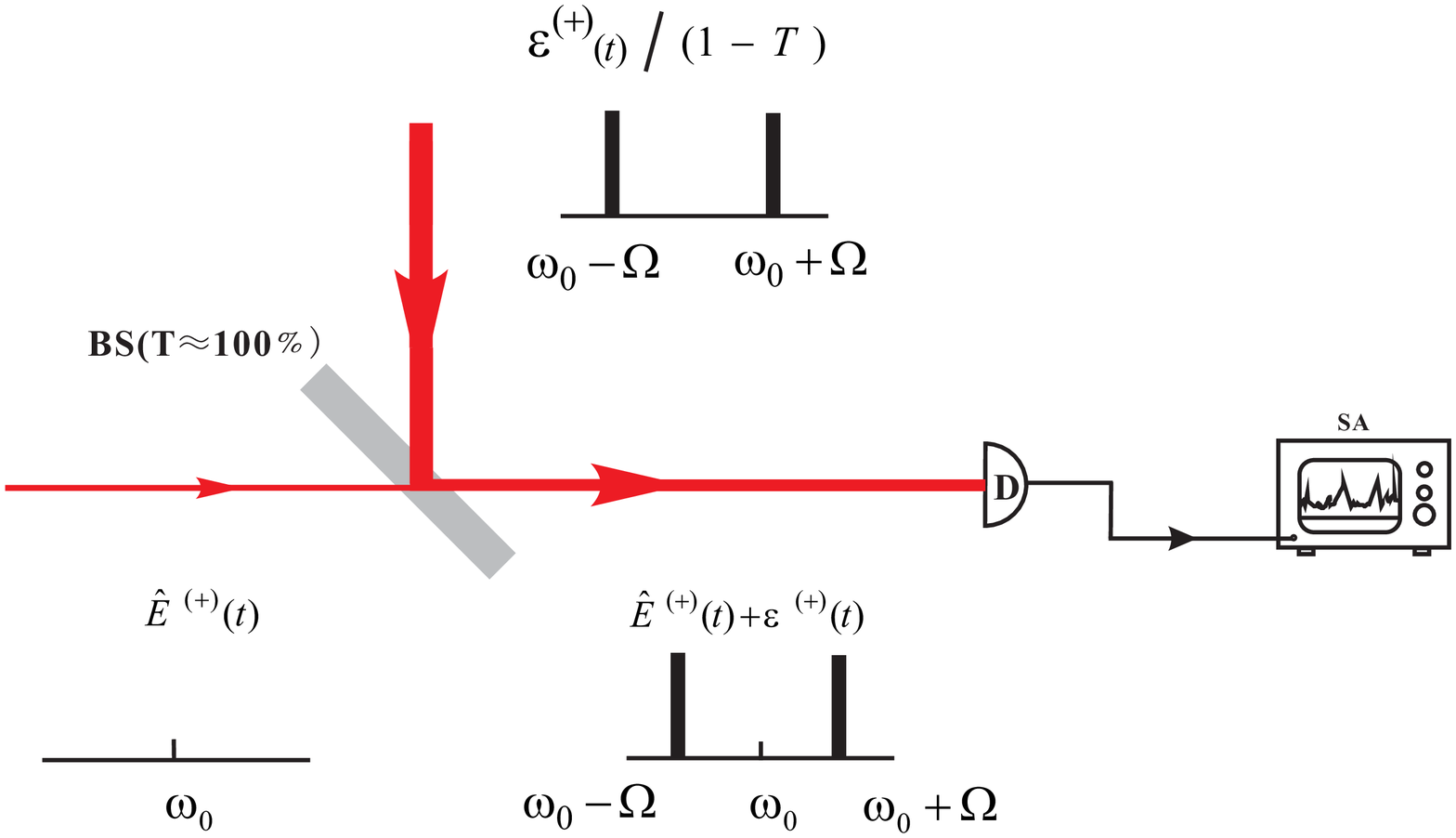}%
\caption{\label{fig:scheme}(color online) Schematic for heterodyne detection of a sub-shot-noise optical signal. The optical field $\hat{E}^{(+)}(t)=\hat{a}e^{-i\omega_0 t}$ to be detected is combined with two local oscillators, $\mathscr{E}_1^{(+)}(t)=\mathscr{E}e^{-i(\omega_0+\Omega) t+i\phi_1}$ and $\mathscr{E}_2^{(+)}(t)=\mathscr{E}e^{-i(\omega_0-\Omega) t+i\phi_2}$, at a beamsplitter with near 100\% transmissivity. $\mathscr{E}^{(+)}(t)\equiv\mathscr{E}_1^{(+)}(t)+\mathscr{E}_2^{(+)}(t)$ stands for the superposition of the two oscillators. Even if the oscillators are attenuated when reflected by the beamsplitter, we still assume that they are much stronger with equal amplitude $\mathscr{E}$ than the detected field. In the frequency domain, the oscillators at $\omega_0\pm\Omega$ are symmetrically located around the detected field at $\omega_0$. $\phi_{1,2} $ are the global phases of the oscillators and $\beta$ is determined by the squeezed quadrature of the detected optical mode. BS: beamsplitter. D: photodetector. SA: spectral analyzer.}
\end{figure}

The balanced-heterodyne scheme manifests itself by utilizing two local oscillators with equal strength that are symmetrically located in frequency spectrum around the optical signal to be measured, as depicted in Fig.~\ref{fig:scheme} with detailed configurations to be described in the next section. Section III is devoted to calculating the spectral density of the photocurrent fluctuations produced by the detector. Unsurprisingly, our calculations show that the detected noise level of the optical signal is phase sensitive, i.e., the noise level depends on some relative phase of light, a feature also shared by homodyne schemes. For some given relative phase, observation of a sub-shot-noise signal is possible if the optical signal is carried in the squeezed quadrature of the field. On the other hand, if the signal is carried in the anti-squeezed quadrature, it must be manipulated to be carried in the squeezed quadrature, using for example a single-ended cavity \cite{galatola1991, kimble2001}, before heterodyne detection is performed. Moreover, we discuss how the additional heterodyne noise can be ruled out in balanced heterodyne detection. What succeeds in Section IV is of how to lock the relative phase of light for balanced heterodyning. The locking technique is quantitatively analyzed to further show the practical feasibility of the studied scheme. Finally, we discuss the potential application of the balanced-heterodyne scheme for practical gravitational-wave observation, followed by a short summary as conclusions to this work by the end with acknowledgements.

\section{Heterodyne detection scheme}
\begin{figure}[B]
\includegraphics[scale=0.5]{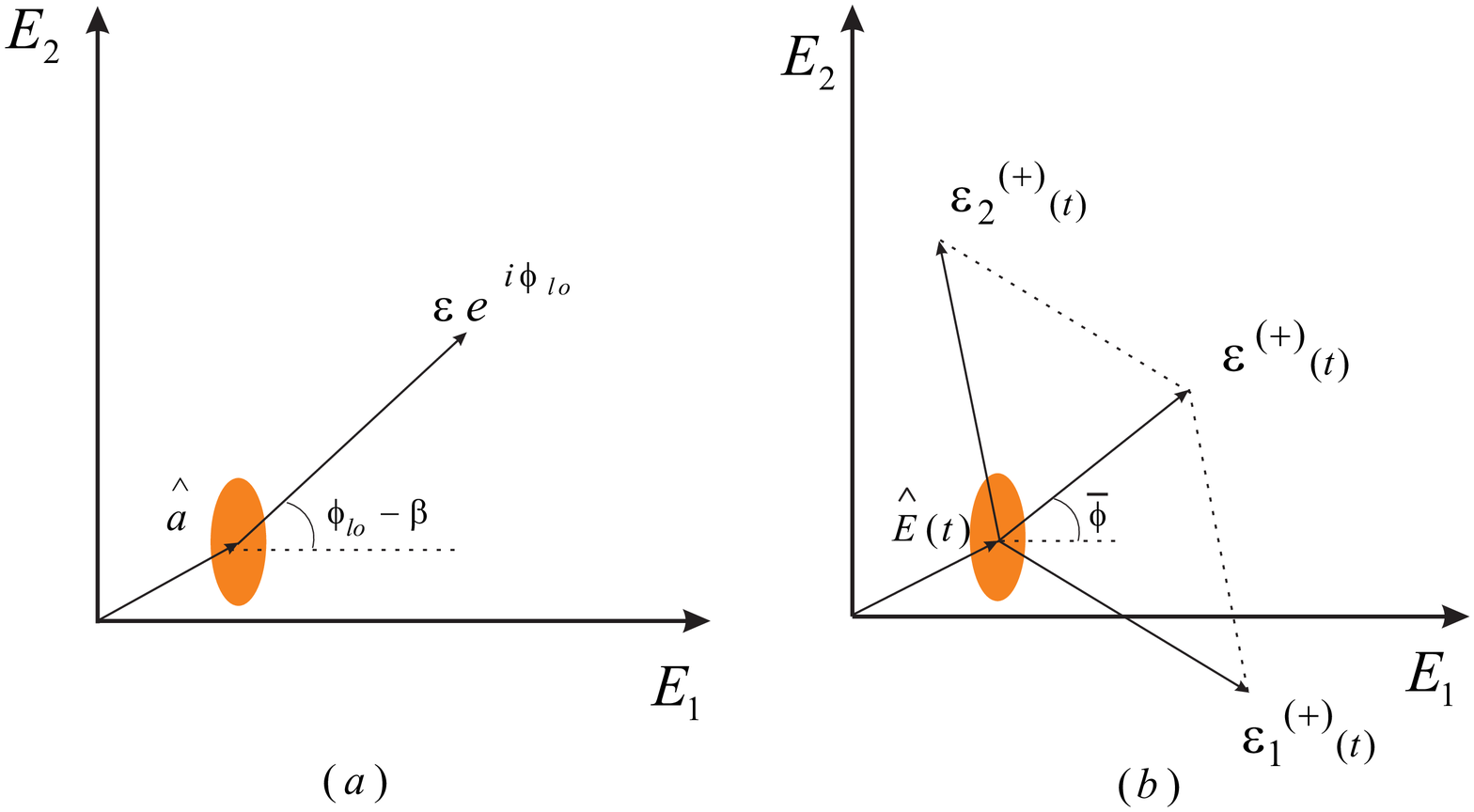}
\caption{\label{fig:explain}(color online) Illustration of the relative orientation of the squeezed quadrature of the detected field and the local oscillator(s) that can be experimentally controlled by some relative phase of light. (a) Conventional homodyne scheme, where controlling the relative phase, $\phi_{lo}-\beta$, of the oscillator $\mathscr{E}e^{i\phi_{lo}}$ with respect to the reference determined by the squeezed quadrature of the field $\hat{a}$ selects which quadrature to be measured. The axis of the squeezed quadrature is parallel to the direction of the oscillator in phase space whenever $\phi_{lo}-\beta=k\pi$ ($k$ any integer). (b) Heterodyne scheme with dual local oscillators $\mathscr{E}_{1}^{(+)}(t)$ and $\mathscr{E}_{2}^{(+)}(t)$, in which the superposition $\mathscr{E}^{(+)}(t)= \mathscr{E}_{1}^{(+)}(t) + \mathscr{E}_{2}^{(+)}(t)$ of the two oscillators remains in a line with a fixed azimuthal angle relative to the squeezed quadrature of the detected field $\hat{E}(t)$. Through controlling the relative phase $\bar{\phi}=(\phi_1+\phi_2)/2-\beta$ defined by $\mathscr{E}^{(+)}(t)$ and $\hat{E}(t)$, one may select which quadrature of $\hat{E}(t)$ to be measured. The direction of the superposed field is parallel to the axis of the measured quadrature, even though each of the oscillators rotates at frequency $\Omega$ clockwise or counter-clockwise with respect to the detected field. See Fig. \ref{fig:scheme} for the definition of $\phi_{1,2}$ and $\beta$.
}
\end{figure}

According to the uncertainty principle, when one quadrature of an optical field is squeezed, its canonical conjugate should be anti-squeezed. In the conventional homodyne detection scheme, one may choose to measure either the squeezed quadrature or its conjugate by controlling the phase of the local oscillator relative to some reference defined by the measured optical mode. Changing this relative phase actually orients the direction of the oscillator along that of the detected quadrature in phase space (Fig.~\ref{fig:explain}a). Only when the local oscillator is parallel to the squeezed quadrature, could one observe the most noise-reduction below the shot-noise level in the homodyne detection. In contrast, the heterodyne scheme with single local oscillator has proven unsuitable for squeezing detection \cite{collett1987}, because the relative phase and, hence, the relative orientation in phase space, of the local oscillator with respect to the squeezed quadrature continuously varies with time and therefore is out of experimental control. As a remedy to this, the balanced-heterodyne scheme takes advantage of two local oscillators with equal strength, the superposition of which oscillates but remains pointing in a fixed direction in phase space, selecting the corresponding quadrature to be measured (Fig.~\ref{fig:explain}b). To see this, we may write the quantity of light intensity at the output port of the beamsplitter in the following form (see Fig. \ref{fig:scheme} for reference)
\begin{eqnarray}
\hat{I}(t) &=& [\hat{E}^{(-)}(t)+\mathscr{E}^{(-)}_1(t)+\mathscr{E}^{(-)}_2(t)]\times[\hat{E}^{(+)}(t)+\mathscr{E}^{(+)}_1(t)+\mathscr{E}^{(+)}_2(t)]\nonumber \\
&=&[e^{i\omega_0 t}(\hat{a}^\dagger+\mathscr{E}e^{i\Omega t-i\phi_1}+\mathscr{E}e^{-i\Omega t-i\phi_2})]\nonumber\\
&\times&
[e^{-i\omega_0 t}(\hat{a}+\mathscr{E}e^{-i\Omega t+i\phi_1}+\mathscr{E}e^{i\Omega t+i\phi_2})]\nonumber\\
&=&
\hat{a}^\dagger\hat{a}+2\mathscr{E}\cos{(\Omega t+\delta \phi)}\hat{X}(\bar{\phi)}+4\mathscr{E}^2\cos^2{(\Omega t+\delta \phi)}. \label{eq:photonnumber}
\end{eqnarray}
Here we adopted units in which the light intensity is expressed in photons per second. $\hat{E}^{(+)}(t)=\hat{a}e^{-i\omega_0 t}$, $\mathscr{E}_1^{(+)}(t)=\mathscr{E}e^{-i(\omega_0+\Omega) t+i\phi_1}$, and $\mathscr{E}_2^{(+)}(t)=\mathscr{E}e^{-i(\omega_0-\Omega) t+i\phi_2}$ represent the optical field to be detected and the two oscillators, respectively. And we assumed that the two oscillators are strongly-excited coherent modes with the same amplitude $\mathscr{E}$ ($\mathscr{E}$ a positive real number) and can be approximated by classical fields. $\hat{X}(\bar{\phi})=\hat{X}\cos{\bar{\phi}}+\hat{P}\sin{\bar{\phi}}$, where $\hat{X}\equiv\hat{a}e^{-i\beta}+\hat{a}^{\dagger}e^{i\beta}$ and $\hat{P}\equiv\hat{a}e^{-i(\beta+\pi/2)}+\hat{a}^{\dagger}e^{i(\beta+\pi/2)}$ are two quadrature amplitudes of the optical field $\hat{E}^{(-)}(t)$. $\bar{\phi}\equiv(\phi_1+\phi_2)/2-\beta$, and $\beta$ is an arbitrary phase that is used to define $\hat{X}$ and $\hat{P}$, thereby the major axis of the distribution ellipse of $\hat{X}$ and $\hat{P}$ in phase space points in the direction of the $\hat{X}$ or $\hat{P}$ axis \cite{ou1987}. $\delta\phi\equiv(\phi_2-\phi_1)/2$ is the phase difference between the two oscillators. 

\subsection{Photocurrent fluctuations}
The first term on the right-hand side of Eq.(\ref{eq:photonnumber}) is the DC signal of the detected field, which is weak compared to other terms and, hence, negligible in strong-oscillator approximation. The last term contributes a classical signal that does not fluctuate when the oscillators are in highly coherent states \cite{mandel82}.  
Accordingly, the middle term dominates in the photon fluctuations: 
\begin{equation}
\Delta\hat{I}(t) \approx 2\mathscr{E}\cos{(\Omega t+\delta \phi)}\Delta\hat{X}(\bar{\phi)},\label{eq:photonnoise}
\end{equation}
which shows phase sensitivity on $\bar{\phi}$ and differs from the homodyne case by only a modulation function $\cos{(\Omega t+\delta \phi)}$. In other words, through controlling the relative phase $\bar{\phi}$, one can choose to measure the noise of any quadrature of the detected field in the balanced-heterodyne scheme, which is analogous to the conventional homodyne scheme. Since the modulation function $\cos{(\Omega t+\delta \phi)}$ in Eq. (\ref{eq:photonnoise}) leaves unchanged the relative orientation of the measured quadrature of the detected field and the superposition of the two oscillators (Fig. \ref{fig:explain}b), its only effects are to split the spectral density of the photocurrent fluctuations into two parts and to shift each part along the frequency axis by $+\Omega$ or $-\Omega$, as will be shown in the following.

\subsection{Photoelectrical signal}
At the output of the detector, the photoelectrical signal, with photon fluctuations neglected, is proportional to the light intensity averaged over the initial states of the optical fields. Using Eq. (\ref{eq:photonnumber}), we may calculate the averaged light intensity as
\begin{eqnarray}\label{eq:photonave}
<\hat{I}(t)>&=&<\hat{a}^\dagger\hat{a}>+2\mathscr{E}\cos{(\Omega t+\delta \phi)}<\hat{X}(\bar{\phi})>+4\mathscr{E}^2\cos^2{(\Omega t+\delta \phi)}.
\label{eq:photocurrent}
\end{eqnarray}
Apparently, the heterodyne signal of interest at frequency $\Omega$ has an amplitude $2\mathscr{E}<\hat{X}(\bar{\phi)}>$ that exhibits a dependence on $\bar{\phi}$, in other words, a photoelectrical signal appearing in $<\hat{X}(\bar{\phi})>$ for a given $\bar{\phi}$ may not show up in the conjugate $<\hat{X}(\bar{\phi}+\pi/2)>$.

Consequently, a naive picture that one may imagine is as follows: To take advantage of optical squeezing to achieve maximal enhancement of signal-to-noise ratio in precision measurements, one must find a $\bar{\phi}$, for which the photoelectrical signal at heterodyne frequency $\Omega$ is maximized and the photon fluctuations are suppressed the most at the same time. To put it another way, for sub-shot-noise detection of an optical signal in balanced-heterodyne scheme, the quadrature $\hat{X}(\bar{\phi})$ in which the optical signal is carried must be a squeezed one. True or not, we expect to find the answer in the subsequent section after calculating the spectral density of the photocurrent fluctuations produced at the output of the detector.

\section{Spectral density of photocurrent fluctuations}
We will follow the approach of Ou, Hong and Mandel \cite{ou1987} to calculate the spectral density of the photocurrent fluctuations. Nonetheless, stationary photocurrents can be assumed only at the time scale of ${\Omega}^{-1}$, beyond which the measured spectral density should be treated as an average over measurement time $T$ (see Ref. \cite{collett1987,drummond90} for similar treatments). Usually $T\sim{\Omega_r}^{-1}>>{\Omega}^{-1}$ ($\Omega_r$ the detection bandwidth. To resolve the heterodyne beat note at $\Omega$, one must set $\Omega_r<<\Omega$). In consequence, the spectral density of the photocurrent fluctuations $\Delta J(t)\equiv J(t)-<J(t)>$ reads

\begin{equation}\label{eq:chi}
\chi(\omega)=\frac{1}{T}{\int}^{T}_{0}dt{\int}^{+\infty}_{-\infty}d\tau e^{i\omega \tau}<\Delta J(t) \Delta J(t+\tau)>,
\end{equation}
where $<\Delta J(t) \Delta J(t+\tau)>$ is the auto-correlation function of the photocurrent fluctuations. Supposing that every photoelectron emitted at time $t'$ gives rise to a definite photoelectrical current pulse $j(t-t')$ for $t>t'$, then the total photocurrent is
\begin{equation}\label{eq:currentsum}
J(t)=\sum_i j(t-t_i),
\end{equation}
wherein the sum is taken over the various random-emission times $t_i$. When $t<t_i$, $j(t-t_i)=0$ because no photoelectrical pulses exist yet prior to the emission of electrons. With Eq. (\ref{eq:currentsum}), the auto-correlation function of the photocurrents can be computed as \cite{ou1987,mandel1995}
\begin{eqnarray}
<J(t)J(t+\tau)>&=&\sum_{l,m}<j(t-t_l)j(t+\tau-t_m)>\nonumber \\
&=& \sum_{l}<j(t-t_l)j(t+\tau-t_l)>+\sum_{l\neq m}<j(t-t_l)j(t+\tau-t_m)>\nonumber\\
&=& \sum_{l}j(t-t_l)j(t+\tau-t_l)P_1(t_l)\Delta t_l \nonumber \\
&&+\sum_{l\neq m}j(t-t_l)j(t+\tau-t_m)P_2(t_l,t_m)\Delta t_l \Delta t_m ,
\label{eq:curcorr}
\end{eqnarray}
where $P_1(t)\Delta t$ and $P_2(t,t')\Delta t \Delta t'$ are respectively the probability of photodetection registered at time $t$ within time interval $\Delta t$ and joint probability that two photodetections are enrolled at $t$ within $\Delta t$ and at $t'$ within $\Delta t'$, respectively. The first term is attributable to the shot noise of the photoelectrical currents, while the second term depends on the fluctuation nature of the optical fields at the photoreceiver. These probabilities are related to the light intensity as \cite{glauber1963a,glauber1963b}
\begin{eqnarray}
P_1(t)\Delta t &=& \eta <\hat{I}(t)> \Delta t\nonumber \\
P_2(t,t')\Delta t \Delta t' &=& \eta^2<\mathscr{T}:\hat{I}(t)\hat{I}(t'):>\Delta t \Delta t' ,
\label{eq:prob}
\end{eqnarray}
in which $\eta$ is a parameter characterizing the response of the detector to the incident light. The symbol $\mathscr{T}: :$ stands for time- and normal-ordering of the field operators. Converting the summation into integrand in Eq. (\ref{eq:curcorr}) and using Eq. (\ref{eq:prob}), one arrives at 
\begin{eqnarray}
<\Delta J(t)\Delta J(t+\tau)>&=&\eta \int^{\infty}_0 dt'<\hat{I}(t-t')>j(t')j(t'+\tau)\nonumber\\
& & +\eta^2 \int\!\!\!\int_0^{\infty}dt'dt''\lambda(t-t',\tau+t'-t'')j(t')j(t''),
\label{eq:intvarcorr}
\end{eqnarray}
where introduced is the correlation function of light-intensity fluctuations
\begin{equation}
\lambda(t,\iota)=<\mathscr{T}:\Delta \hat{I}(t)\Delta \hat{I}(t+\iota):>.
\end{equation}
One should recall that $<\hat{I}(t)>$ in heterodyne detection is not time-independent, in contrast to the homodyne case where $<\hat{I}(t)>$ can be assumed to be stationary. The calculation procedure of the intensity correlation function $\lambda(t,\iota)$ is trivial, and the treatment is resemblant to that based on the approach of Ou, Hong and Mandel \cite{ou1987}. The key idea is to establish a connection between the spectral density of the photocurrent fluctuations and the normally ordered, time-ordered correlation functions of the quadrature fluctuations of the optical field detected. 

For the heterodyne configuration as in Fig.~\ref{fig:scheme}, the correlation function of intensity fluctuations is found to be different from that for the traditional homodyne configuration (see the appendix for detailed calculations):
\begin{eqnarray}
\lambda(t,\iota)&=&<\mathscr{T}:\Delta \hat{I}(t)\Delta \hat{I}(t+\iota):> \nonumber \\
&=& \mathscr{E}^2 \{\Gamma^{(1,1)}(\iota)\left[e^{i\Omega \iota}+e^{-i\Omega \iota}+e^{-i\Omega(2t+\iota)-i2\delta \phi }+e^{i\Omega(2t+\iota)+2i\delta \phi}\right]\nonumber\\
&+&\Gamma^{(2,0)}(\iota)\left[e^{i\Omega \iota+i(\phi_1+\phi_2)}+e^{-i\Omega \iota+i(\phi_1+\phi_2)}+e^{-i\Omega(2t+\iota)+2i\phi_1}+e^{i\Omega(2t+\iota)+2i\phi_2}\right]+c.c. \}+ O(\mathscr{E}),\nonumber \\ \label{eq:lambda}
\end{eqnarray}
wherein $\Gamma^{(1,1)}(\iota)\equiv <\Delta \hat{E}^{(-)}(t)\Delta \hat{E}^{(+)}(t+\iota)>e^{i\omega_0\iota}$ and $\Gamma^{(2,0)}(\iota)\equiv <\Delta \hat{E}^{(-)}(t)\Delta \hat{E}^{(-)}(t+\iota)>e^{-i\omega_0(2t+\iota)}$. Eq. (\ref{eq:lambda}) obviously contains terms that vary with a temporal period of $(2\Omega)^{-1}$. Because the observed spectral density of the photocurrent fluctuations are averaged over a time period $T\propto \Omega_r^{-1}>>\Omega^{-1}$, these time-dependent terms will be ruled out of Eq. (\ref{eq:chi}) by the temporal integrand and not show up in the spectral density. For this reason, we will keep only the time-independent terms in Eq. (\ref{eq:lambda}) and replace the correlation function $\lambda(t,\iota)$ with a $t$-independent function
\begin{eqnarray}\label{eq:lambda'}
\lambda'(\tau)&\approx&2\mathscr{E}^2\cos{(\Omega\tau)} \{\Gamma^{(1,1)}(\tau)+\Gamma^{(2,0)}(\tau)e^{i(\phi_1+\phi_2)}+c.c. \},
\end{eqnarray}
in the succeeding calculations.  Here the strong-oscillator approximation is utilized such that the terms in $\mathscr{E}^2$ in Eq. (\ref{eq:lambda}) dominate over all others, which have henceforth been discarded.
 Using Eq. (\ref{eq:photocurrent}), similar treatment can be applied the $<\hat{I}(t)>$ term on the right-hand side of Eq. (\ref{eq:intvarcorr}), resulting in $\frac{1}{T}\int_0^T dt <\hat{I}(t)>=2\mathscr{E}^2$. Plugging Eqs. (\ref{eq:intvarcorr}) and (\ref{eq:lambda}) into Eq. (\ref{eq:chi}) with these manipulations leads to
\begin{equation}\label{eq:chinew}
\chi(\omega)={\int}^{+\infty}_{-\infty}d\tau e^{i\omega \tau}
\left[2\mathscr{E}^2\eta\int^{\infty}_0dt'j(t')j(t'+\tau)+\eta^2\int\!\!\!\int^{\infty}_0dt'dt''j(t')j(t'')\lambda'(\tau+t'-t'')\right].
\end{equation}
In comparison to the spectral density of the photocurrent fluctuations in conventional homodyne scheme \cite{ou1987, mandel1995}, Eq. (\ref{eq:chinew}) differs by a global factor of two and a $\cos{(\Omega \tau)}$ function (see Eq. (\ref{eq:lambda'})) in the second term. A factor of two means that the spectral density in our heterodyne scheme is 3dB higher, which is obviously due to the usage of two local oscillators. As for the function $\cos{(\Omega \tau)}$, as one will see, it plays a crucial role in the heterodyne scheme in splitting squeezing spectrum into two parts and shifting their centers off the carrier frequency, leading to the ``traditional heterodyne noise'' for certain experimental parameters.

The spectral density given by Eq. (\ref{eq:chinew}) can be easily related to the time-ordered, normally ordered second-order correlation functions of the quadrature fluctuations of the detected field through \cite{mandel1995}
\begin{eqnarray}
\mbox{Re}\Gamma^{(1,1)}(\tau)&=&\left[\Gamma_{11}(\tau)+\Gamma_{22}(\tau)\right]/4 \nonumber \\
\mbox{Im}\Gamma^{(1,1)}(\tau)&=&\left[\Gamma_{12}(\tau)-\Gamma_{21}(\tau)\right]/4 \nonumber \\
\mbox{Re}\Gamma^{(2,0)}(\tau)e^{2i\beta}&=&\left[\Gamma_{11}(\tau)-\Gamma_{22}(\tau)\right]/4 \nonumber \\
\mbox{Im}\Gamma^{(2,0)}(\tau)e^{2i\beta}&=&-\left[\Gamma_{12}(\tau)+\Gamma_{21}(\tau)\right]/4,
\label{eq:biggamma}
\end{eqnarray}
where $\Gamma_{mn}(\tau)\equiv<\mathscr{T}:\Delta\hat{E}_m(t)\Delta\hat{E}_n(t+\tau):>,(m,n=1,2)$, and $\hat{E}_1(t)$, $\hat{E}_2(t)$ are the quadrature operators defined as
\begin{eqnarray}
\hat{E}_1(t)&=&\hat{E}^{(+)}(t)e^{i(\omega_0t-\beta)}+\hat{E}^{(-)}(t)e^{-i(\omega_0t-\beta)}\nonumber \\
\hat{E}_2(t)&=&\hat{E}^{(+)}(t)e^{i(\omega_0t-\beta-\pi/2)}+\hat{E}^{(-)}(t)e^{-i(\omega_0t-\beta-\pi/2)}.
\end{eqnarray}
$\beta$ is the arbitrary phase associated with the field quadrature $\hat{E}^{(-)}(t)$, and $\Delta\hat{E}\equiv\hat{E}-<\hat{E}>$. After substituting Eqs. (\ref{eq:biggamma}) into Eq. (\ref{eq:lambda'}), one achieves
\begin{eqnarray}\label{eq:nlambda'}
\lambda'(\tau)&\approx&\mathscr{E}^2\cos{(\Omega\tau)} \{\Gamma_{11}(\tau)(1+\cos{2\bar{\phi}})+\Gamma_{22}(\tau)(1-\cos{2\bar{\phi}})+\left[\Gamma_{12}(\tau)+\Gamma_{21}(\tau)\right]\sin{2\bar{\phi}}\},\nonumber \\
\end{eqnarray}
$\bar{\phi}=(\phi_1+\phi_2)/2-\beta$ again. Let $K(\omega)$ be the Fourier transform of the photoelectrical current pulse $j(t)$,
\begin{equation}\label{eq:komega}
K(\omega)= \int_0^{\infty}d\tau j(\tau)e^{i\omega\tau},
\end{equation}
which may be interpreted as the frequency response of the detector, and let $\Phi_{mn}(\omega)$ the Fourier transform of the correlation functions $\Gamma_{mn}(\tau)$ of the field quadrature fluctuations:
\begin{equation}\label{eq:Lomega}
\Phi_{mn}(\omega)= \int_{-\infty}^{+\infty}d\tau \Gamma_{mn}(\tau)e^{i\omega\tau} (m,n=1,2).
\end{equation}
With the help of Eqs. (\ref{eq:nlambda'}), (\ref{eq:komega}), and (\ref{eq:Lomega}), one can readily rewrite the spectral density of the photocurrent fluctuations Eq. (\ref{eq:chinew}) as
\begin{eqnarray}
\chi(\omega)&\approx&\eta\mathscr{E}^2|K(\omega)|^2\{2+\nonumber\\
&+&(\eta/2)\left[\Phi_{11}(\omega+\Omega)+\Phi_{11}(\omega-\Omega)\right](1+\cos{2\bar{\phi}})\nonumber \\
&+&(\eta/2)\left[\Phi_{22}(\omega+\Omega)+\Phi_{22}(\omega-\Omega)\right](1-\cos{2\bar{\phi}})\nonumber\\
&+&(\eta/2)\left[\Phi_{12}(\omega+\Omega)+\Phi_{21}(\omega+\Omega)+\Phi_{12}(\omega-\Omega)+\Phi_{21}(\omega-\Omega)\right]\sin{2\bar{\phi}} \}. \label{eq:chiomega}
\end{eqnarray}
Particularly, in the special case that $\bar{\phi}=k\pi$ ($k$ any integer),
\begin{equation}
\chi(\omega)\approx 2\eta\mathscr{E}^2|K(\omega)|^2\{1+(\eta/2)\left[\Phi_{11}(\omega+\Omega)+\Phi_{11}(\omega-\Omega)\right]\}. \label{eq:chiomega1}
\end{equation}
It follows that, if $\Phi_{11}<0$, the measured spectral density $\chi(\omega)$ is to fall below the vacuum level for $\bar{\phi}=k\pi$, which is similar to the case of the conventional homodyne detection \cite{ou1987, mandel1995}. When $\bar{\phi}=k\pi\pm\pi/2$ ($k$ any integer),
\begin{equation}
\chi(\omega)\approx 2\eta\mathscr{E}^2|K(\omega)|^2\{1+(\eta/2)\left[\Phi_{22}(\omega+\Omega)+\Phi_{22}(\omega-\Omega)\right]\}. \label{eq:chiomega2}
\end{equation}
Again, akin to the homodyne scheme, the spectral density will be lower than the vacuum level for 
$\bar{\phi}=k\pi\pm\pi/2$, if $\Phi_{22}<0$. Whether $\Phi_{11}$ or $\Phi_{22}$ is to be observed is conditioned on the specific values of $\bar{\phi}$, confirming the intuitive conclusion drawn upon Eq. (\ref{eq:photonnoise}) together with Eq. (\ref{eq:photonave}): To achieve sub-shot-noise detection of an optical signal with the best signal-to-noise ratio enhancement in the balanced-heterodyne scheme, the optical signal must appear in the carrier's quadrature $<\hat{X}(\bar{\phi})>$ that is squeezed. 

Let one consider an optical signal carried in the amplitude of a field. The heterodyne signal, proportional to $<\hat{X}(\bar{\phi})>$ according to Eq. (\ref{eq:photocurrent}), at frequency $\Omega$ is maximized when $\bar{\phi}=k\pi$ ($k$ integer) because it is carried in the amplitude of the field. In this case, noise reduction below the shot noise level is available in the photocurrent only if $\Phi_{11}<0$, as is shown by Eq. (\ref{eq:chiomega1}), meaning quadrature-amplitude squeezing. On the contrary, if it is the quadrature phase of the incident field that is squeezed, i.e., $\Phi_{22}<0$, one cannot observe amplitude-maximized signal with noise reduction, since maximized heterodyne signal demands $\bar{\phi}=k\pi$ and, on the other hand, to observe $\Phi_{22}<0$ requires $\bar{\phi}=k\pi\pm\pi/2$. Consequently, in the case of $\Phi_{22}<0$, before the balanced-heterodyne scheme is applied to yield a sub-shot-noise signal, one needs to rotate the axis of the squeezed quadrature by an angle of $90^{\circ}$ in phase space, for instance, with a single-ended cavity \cite{galatola1991}, such that a quadrature-phase squeezing $\Phi_{22}<0$ is converted into a quadrature-amplitude squeezing $\Phi_{11}<0$. 

\begin{figure}[B]
\includegraphics[scale=0.35]{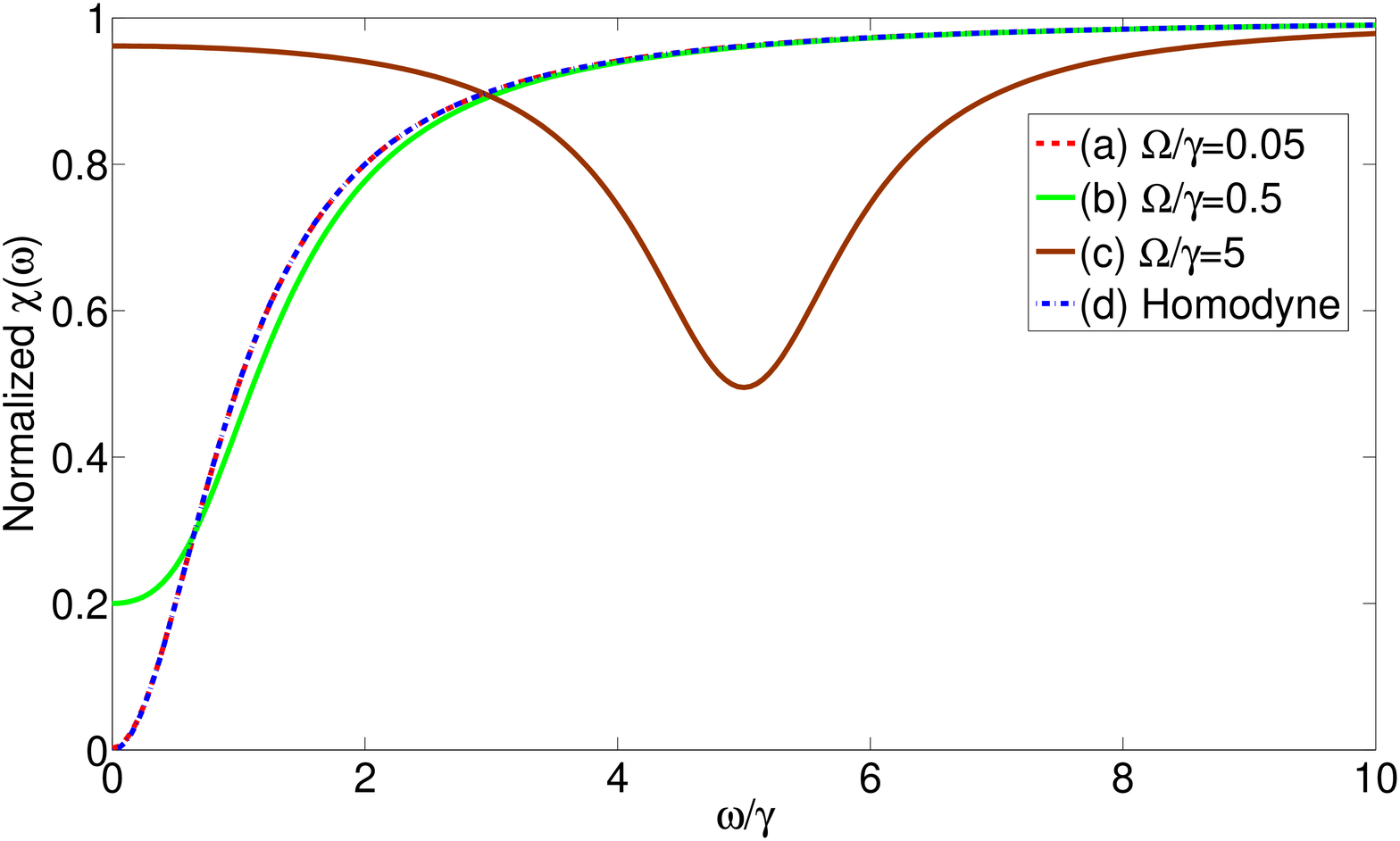}%
\caption{\label{fig:simulation}(color online) Numerical simulations for noise reduction in balanced-heterodyne detection (using Eq. (\ref{eq:simu})) and homodyne detection of sub-shot-noise optical signals, assuming $\gamma/2\epsilon=1$ for simplicity. Normalized $\chi(\omega)$ refers to $\chi(\omega)/(2\eta\mathscr{E}^2|K(\omega)|^2)$, see Eq. (\ref{eq:simu}) for reference. (a) Heterodyning with $\Omega/\gamma=0.05$. Almost perfect squeezing is present in the photocurrent fluctuations, indicating the absence of the additional heterodyne noise present in other heterodyne schemes \cite{niebauer1991, meers1991, buonanno2003} due to vacuum fluctuations in frequency bands that are twice the heterodyne frequency $\Omega$ away from the carrier. (b) Heterodyning with $\Omega/\gamma = 0.5$. The degree of squeezing shows a tendency of degradation due to increased quantum noise at $\omega_0\pm 2\Omega$. (c) Heterodyning with $\Omega/\gamma=5$. The greatest degree of squeezing is only 3dB (50\% noise reduction), showing vacuum fluctuations at $\omega_0\pm 2\Omega$. (d) Homodyne detection with the same squeezed-vacuum mode detected as in (a)-(c).
}
\end{figure}

On the other hand, Eqs. (\ref{eq:chiomega1}) and (\ref{eq:chiomega2}) both entail two equally-split parts of $\Phi_{ii}$($1=1,2$), each shifting away from the original center by $\pm\Omega$, as a direct consequence of the existence of $\cos(\Omega\tau)$ in Eq. (\ref{eq:lambda'}), which is nothing but our enthusiastic argument previously made on Eq. (\ref{eq:photonnoise}). This will cause an issue in practice when applying balanced-heterodyne scheme in precision measurements: The spectrum of squeezing $\Phi_{11}(\omega)$ or $\Phi_{22}(\omega)$ can put a limit to the degree of noise reduction in balanced-heterodyne detection. For a conceptual illustration, let one consider a squeezed-vacuum field generated from an optical parametric down-converter in a cavity. Suppose that this squeezed field is to be heterodyned with two local oscillators in the configuration as depicted in Fig.~\ref{fig:scheme}. The spectrum of squeezing at the output of the cavity is \cite{ou1987}
\begin{equation}
\Phi_{11}=-\frac{2}{\eta}\frac{\epsilon\gamma}{(\gamma/2+\epsilon)^2+\omega^2},
\end{equation}
where $\gamma\equiv(1-\mathscr{R})c/l$ ($\mathscr{R}$ is the cavity mirror reflectivity, $c$ the speed of light in vacuum, and $l$ the cavity length) is the cavity damping rate and $\epsilon$ is a measure of the effective pump intensity of the down-converter. Obviously, $\gamma/2\epsilon$ is a key parameter that determines the spectrum of squeezing. To achieve sub-shot-noise detection with signal-to-noise ratio improved the most, one must control the relative phase $\bar{\phi}=k\pi$, so that the spectral density 
\begin{equation}\label{eq:simu}
\chi(\omega)\approx 2\eta\mathscr{E}^2|K(\omega)|^2\left[1-\frac{\epsilon\gamma}{(\gamma/2+\epsilon)^2+(\omega+\Omega)^2}-\frac{\epsilon\gamma}{(\gamma/2+\epsilon)^2+(\omega-\Omega)^2}
\right].
\end{equation}
Numerical simulations based on this equation (Fig.~\ref{fig:simulation}) show that, to fully take advantage of the degree of squeezing of the incident signal, the heterodyne frequency must satisfy $\Omega<<\gamma$. This is the condition under which ``the additional heterodyne noise'' can be completely removed in the studied scheme, at least, in principle. This may find interesting applications in precision measurements, such as in the observation of gravitational waves. In the limit of $\Omega>>\gamma$, one can obtain a noise reduction by 3dB at most, because of the apparent fact that $\Phi_{11}(\omega+\Omega)$ makes no contribution to noise squeezing at frequencies where $\Phi_{11}(\omega-\Omega)$ does its best, the physics behind which is the vacuum fluctuations twice the heterodyne frequency away from the carrier, an important subject of previous works \cite{gea1987, niebauer1991, meers1991, buonanno2003}.

\section{Phase locking technique}
\begin{figure}[B]
\includegraphics[scale=0.56]{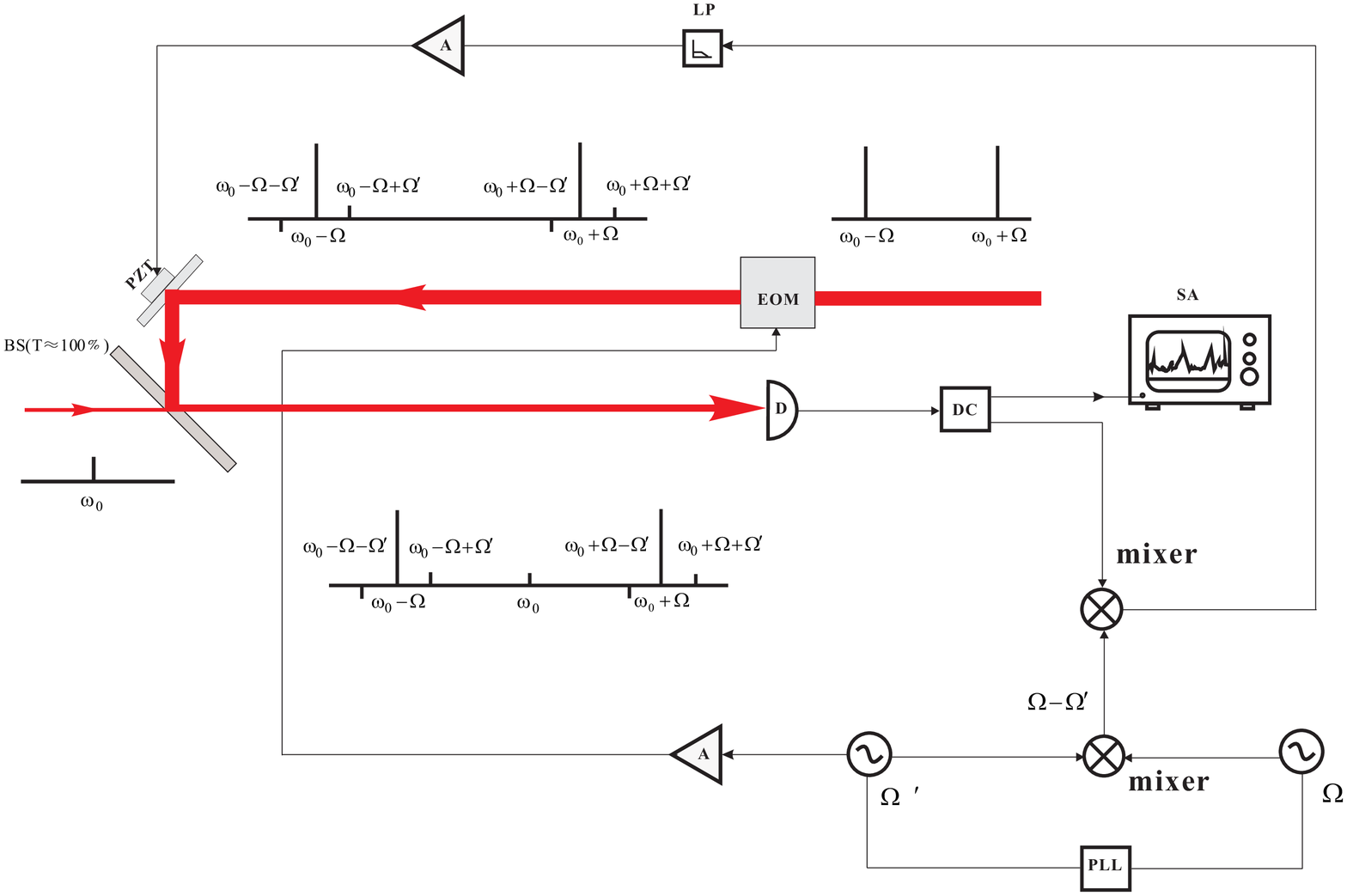}
\caption{\label{fig:locking}(color online) Phase locking scheme for balanced-heterodyne detection of sub-shot-noise optical signals. The two oscillators are phase-modulated by the same modulator at a RF frequency $\Omega'<\Omega$ with a modulation depth $\theta$. The error signal is fed into a piezo-electrical transducer to change the global phase of the oscillators, after the photoelectrical signal is demodulated at frequency $\Omega-\Omega'$. Obviously, the two RF signals at $\Omega$ and $\Omega'$ must be phase-locked to each other. The control part for the relative phase of the two oscillators is omitted in the figure, albeit also important as explained in the text. EOM: electro-optical modulator. PZT: piezo-electrical transducer. BS: beamsplitter. D: photodetector. DC: electrical directional coupler. SA: spectral analyzer. LP: loop filter. A: amplifiers. $\Omega$ and $\Omega'$: signal generators. PLL: electrical phase-locking loop.
}
\end{figure}

To enforce the balanced-heterodyne scheme for sub-shot-noise detection, the optical signal to be measured should appear in the squeezed quadrature of the carrier field, as discussed earlier. Meanwhile, the relative phase $\bar{\phi}=(\phi_1+\phi_2)/2-\beta$ must be well controlled at some values for maximal $<\hat{X}(\bar{\phi})>$ according to Eq. (\ref{eq:photonave}), requiring an electrical locking technique for this scheme. A well known technique of quantum noise locking \cite{mckenzie2005} proposed for homodyne scheme apparently is not applicable to the balanced-heterodyne scheme entailing two local oscillators. We propose to phase-modulate the two oscillators simultaneously with the same modulator and utilize the interference of the detected optical signal with two of the four sidebands of the oscillators, see Fig.~\ref{fig:locking}. The beat notes of the detected signal and those modulation-created sidebands are to be demodulated, then low-pass-filtered by a loop-filter to obtain an error signal for the locking loop. This phase-locking scheme is actually a version of the coherent-modulation-locking technique, whose stability is much better than the noise locking scheme for homodyne detection \cite{mckenzie2005}.

At the output of the photodetector, the photocurrent is \cite{mandel1995}
\begin{eqnarray}\label{eq:jt}
<J(t)>&=&\int_{-\infty}^{+\infty}dt' P_1(t') j(t-t')\nonumber\\
&=& \int_{-\infty}^{+\infty}dt' \Big[\eta <\hat{I}(t-t')>\Big] j(t')\nonumber\\
&\approx&\eta q <\hat{I}(t)>,
\end{eqnarray}
in which $q\equiv\int_{-\infty}^{+\infty}dt'j(t')$ is the total electrical charge delivered by the current pulse resulting from one photoelectron. Here one assumed that the electrical current pulse is much shorter than the oscillation period of $<\hat{I}(t)>$, in another word, $<\hat{I}(t)>$ varies slowly compared to electrical pulse $j(t)$, which is usually the case in practice. Under this assumption, $<\hat{I}(t)>$ may be considered approximately constant as long as the electrical pulse lasts and does not contribute to the integral in Eq. (\ref{eq:jt}). 

After passing the same electro-optical modulator, the oscillator fields become
\begin{equation}
\mathscr{E}^{(+)}_{1,2}(t)=\mathscr{E}e^{-i(\omega_0\pm\Omega) t+i(\phi_{1,2}+\theta\sin{\Omega't})},
\end{equation}
wherein $\Omega'<\Omega$ is the phase-modulation frequency and $\theta$ the modulation depth. Then
\begin{eqnarray}
<J(t)>&\approx& \eta q <[\hat{E}^{(-)}(t)+\mathscr{E}^{(-)}_1(t)+\mathscr{E}^{(-)}_2(t)]\times[\hat{E}^{(+)}(t)+\mathscr{E}^{(+)}_1(t)+\mathscr{E}^{(+)}_2(t)]>\nonumber \\
&=&\eta q <(\hat{a}^\dagger +\mathscr{E}e^{i\Omega t-i\phi_1(t)}+\mathscr{E}e^{-i\Omega t-i\phi_2(t)})\times
(\hat{a}+\mathscr{E}e^{-i\Omega t+i\phi_1(t)}+\mathscr{E}e^{i\Omega t+i\phi_2(t)})>\nonumber\\
&\approx&\eta q \mathscr{E} J_1(\theta)<\hat{a}>e^{-i\phi_1}e^{i(\Omega-\Omega')t}-\eta q\mathscr{E} J_1(\theta)<\hat{a}>e^{-i\phi_2}e^{-i(\Omega-\Omega')t}\nonumber\\
&+& \eta q\mathscr{E} J_1(\theta)<\hat{a}^{\dagger}>e^{i\phi_1}e^{-i(\Omega-\Omega')t}-\eta q\mathscr{E} J_1(\theta)<\hat{a}^{\dagger}>e^{i\phi_2}e^{i(\Omega-\Omega')t}+...,
\end{eqnarray}
where an appropriate modulation depth is assumed so that $e^{i\theta\sin{\Omega' t}}\approx J_0(\theta)+J_1(\theta)e^{i\Omega't}-J_1(\theta)e^{-i\Omega't}$ with $J_{0,1}(\theta)$ the zero- and the first-order Bessel functions. In the last step, explicitly written out are the terms oscillating at frequency $\Omega-\Omega'$ and the reason is that a demodulation signal at this frequency is to be used for error-signal pick up. Hence, the error signal to be demodulated reads
\begin{equation}
S(\bar{\phi})=4\eta q\mathscr{E}J_1(\theta)\Big[\frac{\partial}{\partial \bar{\phi}}<\hat{X}(\bar{\phi})>\Big]\sin{\left[(\Omega-\Omega')t+\delta\phi\right]}.
\end{equation}
The term $\frac{\partial}{\partial \bar{\phi}}<\hat{X}(\bar{\phi})>$ in this equation ensures that $\bar{\phi}$ can be readily locked at any values for which $<\hat{X}(\bar{\phi})>$ is maximal, i.e., the peak of the photoelectrical signal. Nevertheless, one must note that the relative phase $\delta\phi$ of the two oscillators should also be kept steady for the locking scheme to work well.

\section{Balanced heterodyning for gravitational-wave observation}

The existence of the additional heterodyne noise makes the previously-investigated schemes \cite{niebauer1991, meers1991, buonanno2003} less competitive than homodyne schemes in practical applications for precision measurements. However, our analysis on the balanced-heterodyne scheme based on the quantum theory of optical coherence shows the possibility of eliminating this additional noise under certain experimental circumstance. Therefore, the balanced-heterodyne scheme may be a promising choice for photoelectrical readout in gravitational-wave observation experiments. The price that one pays is that the heterodyne frequency $\Omega$ is not free to choose. Instead, the choice of $\Omega$ must guarantee that the degree of squeezing at optical frequencies $\omega_0\pm 2\Omega$ is more or less the same as at $\omega_0$ (Fig. \ref{fig:simulation}).

Since the balanced-heterodyne scheme makes use of dual local oscillators with equal strength, the unbalanced Schnupp sidebands generated by phase modulation in Advanced LIGO \cite{buonanno2003} cannot be used as the balanced local-oscillators. However, one might pick up part of the light from the bright port of the interferometer to serve as oscillators for balanced heterodyning after some appropriate manipulations. If this is the choice, the proposed phase-locking technique in the preceding section may be exploited to control the fluctuations of the involved phases of light.

\section{Conclusions}
Although homodyne detection scheme has been widely used to measure the quantum noise of light below the shot-noise level in the field of quantum information processing, it can hardly be exploited to detect slowly-varying sub-shot-noise optical signals due to the intrinsic electrical noise of the photodetector at low frequencies. To fulfill the need of noise reduction in precision experiments where the shot noise of light is becoming a substantial restrictive factor for further sensitivity improvement, we have investigated the balanced-heterodyne detection scheme, where observation of noise reduction below the shot-noise level is sensitive to some relative phase defined by the signal-carrying optical field and the oscillators, a trait akin to the classical homodyne schemes. An optical signal carried in the squeezed quadrature of the carrier field can be measured with the best signal-to-noise ratio enhancement, whereas that carried by the quadrature conjugate to the squeezed one may be manipulated by a single-ended cavity before a balanced-heterodyne detection is performed. Most importantly, we have shown that, under certain experimental circumstances, the additional heterodyne noise existing in previously-studied schemes may disappear in balanced-heterodyne detection, which may be particularly interesting in applications for gravitational-wave searching. A phase-locking technique has been analyzed to yonder demonstrate the feasibility of the scheme for practical implementation. 

\appendix*
\section{}

According to the definition, the correlation function of light-intensity fluctuations reads
\begin{eqnarray}
\lambda(t,\iota)&=&<\mathscr{T}:\Delta \hat{I}(t)\Delta \hat{I}(t+\iota):> \nonumber \\ 
&=& <\mathscr{T}: \hat{I}(t) \hat{I}(t+\iota):>-<\hat{I}(t)><\hat{I}(t+\iota)>, \label{eq:app1}
\end{eqnarray}
where $\hat{I}(t) = [\hat{E}^{(-)}(t)+\mathscr{E}^{(-)}_1(t)+\mathscr{E}^{(-)}_2(t)]\times[\hat{E}^{(+)}(t)+\mathscr{E}^{(+)}_1(t)+\mathscr{E}^{(+)}_2(t)]$ in conformity with Eq. (\ref{eq:photonnumber}). One may expand the two terms on the right-hand side of Eq. (\ref{eq:app1}) separately as follows:
\begin{eqnarray}
& & <\mathscr{T}: \hat{I}(t) \hat{I}(t+\iota):> \nonumber \\
&=& <[\hat{E}^{(-)}(t)+\mathscr{E}^{(-)}_1(t)+\mathscr{E}^{(-)}_2(t)]\times[\hat{E}^{(-)}(t+\iota)+\mathscr{E}^{(-)}_1(t+\iota)+\mathscr{E}^{(-)}_2(t+\iota)]\nonumber \\
&\times&[\hat{E}^{(+)}(t+\iota)+\mathscr{E}^{(+)}_1(t+\iota)+\mathscr{E}^{(+)}_2(t+\iota)]\times[\hat{E}^{(+)}(t)+\mathscr{E}^{(+)}_1(t)+\mathscr{E}^{(+)}_2(t)]> \nonumber \\
&=& [\mathscr{E}^{(-)}_1(t)+\mathscr{E}^{(-)}_2(t)][\mathscr{E}^{(+)}_1(t)+\mathscr{E}^{(+)}_2(t)][\mathscr{E}^{(-)}_1(t+\iota)+\mathscr{E}^{(-)}_2(t+\iota)][\mathscr{E}^{(+)}_1(t+\iota)+\mathscr{E}^{(+)}_2(t+\iota)] \nonumber \\
&+&[\mathscr{E}^{(+)}_1(t)+\mathscr{E}^{(+)}_2(t)][\mathscr{E}^{(-)}_1(t+\iota)+\mathscr{E}^{(-)}_2(t+\iota)][\mathscr{E}^{(+)}_1(t+\iota)+\mathscr{E}^{(+)}_2(t+\iota)]<\hat{E}^{(-)}(t)> \nonumber \\
&+&[\mathscr{E}^{(-)}_1(t)+\mathscr{E}^{(-)}_2(t)][\mathscr{E}^{(-)}_1(t+\iota)+\mathscr{E}^{(-)}_2(t+\iota)][\mathscr{E}^{(+)}_1(t+\iota)+\mathscr{E}^{(+)}_2(t+\iota)]<\hat{E}^{(+)}(t)> \nonumber \\
&+&[\mathscr{E}^{(-)}_1(t)+\mathscr{E}^{(-)}_2(t)][\mathscr{E}^{(+)}_1(t)+\mathscr{E}^{(+)}_2(t)][\mathscr{E}^{(+)}_1(t+\iota)+\mathscr{E}^{(+)}_2(t+\iota)]<\hat{E}^{(-)}(t+\iota)> \nonumber \\
&+&[\mathscr{E}^{(-)}_1(t)+\mathscr{E}^{(-)}_2(t)][\mathscr{E}^{(+)}_1(t)+\mathscr{E}^{(+)}_2(t)][\mathscr{E}^{(-)}_1(t+\iota)+\mathscr{E}^{(-)}_2(t+\iota)]<\hat{E}^{(+)}(t+\iota)> \nonumber \\
&+&[\mathscr{E}^{(-)}_1(t+\iota)+\mathscr{E}^{(-)}_2(t+\iota)][\mathscr{E}^{(+)}_1(t+\iota)+\mathscr{E}^{(+)}_2(t+\iota)]<\hat{E}^{(-)}(t)\hat{E}^{(+)}(t)> \nonumber \\
&+&[\mathscr{E}^{(-)}_1(t)+\mathscr{E}^{(-)}_2(t)][\mathscr{E}^{(+)}_1(t)+\mathscr{E}^{(+)}_2(t)]<\hat{E}^{(-)}(t+\iota)\hat{E}^{(+)}(t+\iota)> \nonumber \\
&+&[\mathscr{E}^{(+)}_1(t)+\mathscr{E}^{(+)}_2(t)][\mathscr{E}^{(-)}_1(t+\iota)+\mathscr{E}^{(-)}_2(t+\iota)]<\hat{E}^{(-)}(t)\hat{E}^{(+)}(t+\iota)> \nonumber \\
&+&[\mathscr{E}^{(-)}_1(t)+\mathscr{E}^{(-)}_2(t)][\mathscr{E}^{(+)}_1(t+\iota)+\mathscr{E}^{(+)}_2(t+\iota)]<\hat{E}^{(-)}(t+\iota)\hat{E}^{(+)}(t)> \nonumber \\
&+&[\mathscr{E}^{(+)}_1(t)+\mathscr{E}^{(+)}_2(t)][\mathscr{E}^{(+)}_1(t+\iota)+\mathscr{E}^{(+)}_2(t+\iota)]<\hat{E}^{(-)}(t)\hat{E}^{(-)}(t+\iota)> \nonumber \\
&+&[\mathscr{E}^{(-)}_1(t)+\mathscr{E}^{(-)}_2(t)][\mathscr{E}^{(-)}_1(t+\iota)+\mathscr{E}^{(-)}_2(t+\iota)]<\hat{E}^{(+)}(t+\iota)\hat{E}^{(+)}(t)> \nonumber \\
&+&[\mathscr{E}^{(-)}_1(t)+\mathscr{E}^{(-)}_2(t)]<\hat{E}^{(-)}(t+\iota)\hat{E}^{(+)}(t+\iota)\hat{E}^{(+)}(t)> \nonumber \\
&+&[\mathscr{E}^{(+)}_1(t)+\mathscr{E}^{(+)}_2(t)]<\hat{E}^{(-)}(t)\hat{E}^{(-)}(t+\iota)\hat{E}^{(+)}(t+\iota)> \nonumber \\
&+&[\mathscr{E}^{(-)}_1(t+\iota)+\mathscr{E}^{(-)}_2(t+\iota)]<\hat{E}^{(-)}(t)\hat{E}^{(+)}(t+\iota)\hat{E}^{(+)}(t)> \nonumber \\
&+&[\mathscr{E}^{(+)}_1(t+\iota)+\mathscr{E}^{(+)}_2(t+\iota)]<\hat{E}^{(-)}(t)\hat{E}^{(-)}(t+\iota)\hat{E}^{(+)}(t)> \nonumber \\
&+&<\hat{E}^{(-)}(t)\hat{E}^{(-)}(t+\iota)\hat{E}^{(+)}(t+\iota)\hat{E}^{(+)}(t)>, 
\label{eq:app2}\\
\nonumber \\
& & <\hat{I}(t)><\hat{I}(t+\iota)> \nonumber \\
&=& <[\hat{E}^{(-)}(t)+\mathscr{E}^{(-)}_1(t)+\mathscr{E}^{(-)}_2(t)]\times[\hat{E}^{(+)}(t)+\mathscr{E}^{(+)}_1(t)+\mathscr{E}^{(+)}_2(t)]>\nonumber \\
&\times&<[\hat{E}^{(-)}(t+\iota)+\mathscr{E}^{(-)}_1(t+\iota)+\mathscr{E}^{(-)}_2(t+\iota)]\times[\hat{E}^{(+)}(t+\iota)+\mathscr{E}^{(+)}_1(t+\iota)+\mathscr{E}^{(+)}_2(t+\iota)]> \nonumber \\
&=& [\mathscr{E}^{(-)}_1(t)+\mathscr{E}^{(-)}_2(t)][\mathscr{E}^{(+)}_1(t)+\mathscr{E}^{(+)}_2(t)][\mathscr{E}^{(-)}_1(t+\iota)+\mathscr{E}^{(-)}_2(t+\iota)][\mathscr{E}^{(+)}_1(t+\iota)+\mathscr{E}^{(+)}_2(t+\iota)] \nonumber \\
&+&[\mathscr{E}^{(+)}_1(t)+\mathscr{E}^{(+)}_2(t)][\mathscr{E}^{(-)}_1(t+\iota)+\mathscr{E}^{(-)}_2(t+\iota)][\mathscr{E}^{(+)}_1(t+\iota)+\mathscr{E}^{(+)}_2(t+\iota)]<\hat{E}^{(-)}(t)> \nonumber \\
&+&[\mathscr{E}^{(-)}_1(t)+\mathscr{E}^{(-)}_2(t)][\mathscr{E}^{(-)}_1(t+\iota)+\mathscr{E}^{(-)}_2(t+\iota)][\mathscr{E}^{(+)}_1(t+\iota)+\mathscr{E}^{(+)}_2(t+\iota)]<\hat{E}^{(+)}(t)> \nonumber \\
&+&[\mathscr{E}^{(-)}_1(t)+\mathscr{E}^{(-)}_2(t)][\mathscr{E}^{(+)}_1(t)+\mathscr{E}^{(+)}_2(t)][\mathscr{E}^{(+)}_1(t+\iota)+\mathscr{E}^{(+)}_2(t+\iota)]<\hat{E}^{(-)}(t+\iota)> \nonumber \\
&+&[\mathscr{E}^{(-)}_1(t)+\mathscr{E}^{(-)}_2(t)][\mathscr{E}^{(+)}_1(t)+\mathscr{E}^{(+)}_2(t)][\mathscr{E}^{(-)}_1(t+\iota)+\mathscr{E}^{(-)}_2(t+\iota)]<\hat{E}^{(+)}(t+\iota)> \nonumber \\
&+&[\mathscr{E}^{(-)}_1(t+\iota)+\mathscr{E}^{(-)}_2(t+\iota)][\mathscr{E}^{(+)}_1(t+\iota)+\mathscr{E}^{(+)}_2(t+\iota)]<\hat{E}^{(-)}(t)\hat{E}^{(+)}(t)> \nonumber \\
&+&[\mathscr{E}^{(-)}_1(t)+\mathscr{E}^{(-)}_2(t)][\mathscr{E}^{(+)}_1(t)+\mathscr{E}^{(+)}_2(t)]<\hat{E}^{(-)}(t+\iota)\hat{E}^{(+)}(t+\iota)> \nonumber \\
&+&[\mathscr{E}^{(+)}_1(t)+\mathscr{E}^{(+)}_2(t)][\mathscr{E}^{(-)}_1(t+\iota)+\mathscr{E}^{(-)}_2(t+\iota)]<\hat{E}^{(-)}(t)><\hat{E}^{(+)}(t+\iota)> \nonumber \\
&+&[\mathscr{E}^{(-)}_1(t)+\mathscr{E}^{(-)}_2(t)][\mathscr{E}^{(+)}_1(t+\iota)+\mathscr{E}^{(+)}_2(t+\iota)]<\hat{E}^{(-)}(t+\iota)><\hat{E}^{(+)}(t)> \nonumber \\
&+&[\mathscr{E}^{(+)}_1(t)+\mathscr{E}^{(+)}_2(t)][\mathscr{E}^{(+)}_1(t+\iota)+\mathscr{E}^{(+)}_2(t+\iota)]<\hat{E}^{(-)}(t)><\hat{E}^{(-)}(t+\iota)> \nonumber \\
&+&[\mathscr{E}^{(-)}_1(t)+\mathscr{E}^{(-)}_2(t)][\mathscr{E}^{(-)}_1(t+\iota)+\mathscr{E}^{(-)}_2(t+\iota)]<\hat{E}^{(+)}(t+\iota)><\hat{E}^{(+)}(t)> \nonumber \\
&+&[\mathscr{E}^{(-)}_1(t)+\mathscr{E}^{(-)}_2(t)]<\hat{E}^{(+)}(t)><\hat{E}^{(-)}(t+\iota)\hat{E}^{(+)}(t+\iota)> \nonumber \\
&+&[\mathscr{E}^{(+)}_1(t)+\mathscr{E}^{(+)}_2(t)]<\hat{E}^{(-)}(t)><\hat{E}^{(-)}(t+\iota)\hat{E}^{(+)}(t+\iota)> \nonumber \\
&+&[\mathscr{E}^{(-)}_1(t+\iota)+\mathscr{E}^{(-)}_2(t+\iota)]<\hat{E}^{(-)}(t)\hat{E}^{(+)}(t)><\hat{E}^{(+)}(t+\iota)> \nonumber \\
&+&[\mathscr{E}^{(+)}_1(t+\iota)+\mathscr{E}^{(+)}_2(t+\iota)]<\hat{E}^{(-)}(t)\hat{E}^{(+)}(t)><\hat{E}^{(-)}(t+\iota)> \nonumber \\
&+&<\hat{E}^{(-)}(t)\hat{E}^{(+)}(t)><\hat{E}^{(-)}(t+\iota)\hat{E}^{(+)}(t+\iota)>. 
\label{eq:app3}
\end{eqnarray}
Since the seven leading terms in the expansion of $<\mathscr{T}: \hat{I}(t) \hat{I}(t+\iota):>$ are identical to those in $<\hat{I}(t)><\hat{I}(t+\iota)>$, they are canceled out when Eqs. (\ref{eq:app2}) and (\ref{eq:app3}) are plugged into Eq. (\ref{eq:app1}). Therefore,
\begin{eqnarray}
\lambda(t,\iota)&=&[\mathscr{E}^{(+)}_1(t)+\mathscr{E}^{(+)}_2(t)][\mathscr{E}^{(-)}_1(t+\iota)+\mathscr{E}^{(-)}_2(t+\iota)]<\Delta\hat{E}^{(-)}(t)\Delta\hat{E}^{(+)}(t+\iota)> \nonumber \\
&+&[\mathscr{E}^{(-)}_1(t)+\mathscr{E}^{(-)}_2(t)][\mathscr{E}^{(+)}_1(t+\iota)+\mathscr{E}^{(+)}_2(t+\iota)]<\Delta\hat{E}^{(-)}(t+\iota)\Delta\hat{E}^{(+)}(t)> \nonumber \\
&+&[\mathscr{E}^{(+)}_1(t)+\mathscr{E}^{(+)}_2(t)][\mathscr{E}^{(+)}_1(t+\iota)+\mathscr{E}^{(+)}_2(t+\iota)]<\Delta\hat{E}^{(-)}(t)\Delta\hat{E}^{(-)}(t+\iota)> \nonumber \\
&+&[\mathscr{E}^{(-)}_1(t)+\mathscr{E}^{(-)}_2(t)][\mathscr{E}^{(-)}_1(t+\iota)+\mathscr{E}^{(-)}_2(t+\iota)]<\Delta\hat{E}^{(+)}(t+\iota)\Delta\hat{E}^{(+)}(t)> \nonumber \\
&+& O(\mathscr{E}), \label{eq:app4}
\end{eqnarray}
where $O(\mathscr{E})$ represents the terms in $\mathscr{E}$ and those without $\mathscr{E}$ in Eqs. (\ref{eq:app2}) and (\ref{eq:app3}), with $\mathscr{E}$ being the field amplitude of the two oscillators. Substituting $\mathscr{E}_1^{(+)}(t)=\mathscr{E}e^{-i(\omega_0+\Omega) t+i\phi_1}$, and $\mathscr{E}_2^{(+)}(t)=\mathscr{E}e^{-i(\omega_0-\Omega) t+i\phi_2}$ into Eq. (\ref{eq:app4}), one can arrive at Eq. (\ref{eq:lambda}) after some simple mathematical manipulations.

\begin{acknowledgments}
This work is supported by Huazhong University of Science and Technology through the Startup Funding for New Faculty. The authors would like to thank Mr. D.C. He for helping preparing Fig. \ref{fig:scheme}, Fig. \ref{fig:explain}, and Fig. \ref{fig:locking} and Ms. Y. Xiao for Fig. \ref{fig:simulation}.
\end{acknowledgments}

\bibliography{heterodyne-rev}

\end{document}